\documentclass[lettersize,journal]{IEEEtran}
\usepackage{amsmath,amsfonts}
\usepackage{algorithmic}
\usepackage{algorithm}
\usepackage{array}
\usepackage[caption=false,font=normalsize,labelfont=sf,textfont=sf]{subfig}
\usepackage{textcomp}
\usepackage{stfloats}
\usepackage{url}
\usepackage{verbatim}
\usepackage{graphicx}
\usepackage{cite}
\usepackage{tabularx}
\usepackage{booktabs}
\usepackage{colortbl}
\usepackage{xcolor}
\usepackage{enumitem}
\usepackage{multirow}
\usepackage{colortbl}
\usepackage{xcolor}
\usepackage{enumitem}
\usepackage{booktabs}

\usepackage[acronym]{glossaries}
\makeglossaries
\newacronym{isac}{ISAC}{Integrated Sensing and Communication}
\newacronym{6g}{6G}{6th Generation}
\newacronym{5g}{5G}{5th Generation}
\newacronym{llm}{LLM}{Large Language Model}
\newacronym{lwam}{LWAM}{Large Wireless AI Model}
\newacronym{ood}{OOD}{Out-of-distribution}
\newacronym{wlam}{WLAM}{Wireless Large AI Model}
\newacronym{rf}{RF}{Radio Frequency}
\newacronym{lwm}{LWM}{Large Wireless Model}
\newacronym{ssta}{SSTA}{Sparse Spatio-Temporal Attention}
\newacronym{ai}{AI}{Artificial Intelligence}
\newacronym{wmfm}{WMFM}{Wireless Multimodal Foundation Model}
\newacronym{ibn}{IBN}{Intent-Based Networking}
\newacronym{ran}{RAN}{Radio Access Network}
\newacronym{genai}{GenAI}{Generative Artificial Intelligence}
\newacronym{csi}{CSI}{channel state information}
\newacronym{mec}{MEC}{Mobile Edge Computing}
\newacronym{mlp}{MLP}{Multi-Layer Perceptron}
\newacronym{lora}{LoRA}{Low-Rank Adaptation}
\newacronym{mmit}{MM-IT}{Multi-Modal Instruction Tuning}
\newacronym{peft}{PEFT}{Perimeter Efficient Fine-Tuning}
\newacronym{ntn}{NTN}{Non-Terrestrial Networks}
\newacronym{cnn}{CNN}{convolutional neural network}
\newacronym{los}{LoS}{line of sight}
\newacronym{nlos}{NLoS}{non-line of sight}
\newacronym{fm}{FM}{foundation model}
\newacronym{icl}{ICL}{in-context learning}
\newacronym{cot}{CoT}{chain of thought}
\newacronym{sr2}{$S-R^2$}{Signal-to-Reasoning and Response}
\newacronym{pera}{PERA}{Perceive-Reason-Act}

\definecolor{tableheader}{HTML}{B0C4DE} 
\begin{document}

\title{PERA: A Perceive-Reason-Act Interface Bridging Sensing, Cognitive Reasoning, and Trustworthy Agentic Response for 6G}
\author{Mohammad Farzanullah, Melike Erol-Kantarci, ~\IEEEmembership{Fellow,~IEEE} and Lajos Hanzo~\IEEEmembership{Life Fellow} \vspace{-3ex}
\thanks{Mohammad Farzanullah, and Melike Erol-Kantarci are with the School of Electrical Engineering and Computer Science, University of Ottawa, Ottawa, ON K1N 6N5, Canada (e-mail: mfarz086@uottawa.ca; melike.erolkantarci@uottawa.ca).
}

\thanks{Lajos Hanzo is with University of Southampton, U.K (email: lh@ecs.soton.ac.uk).}
}

\maketitle

\begin{abstract}

The realization of next-generation (NG) networks hinges on a fundamental departure from pre-programmed protocol engineering towards a paradigm of self-consciously evolving, autonomous and trusted intelligence. While conventional machine learning (ML) has introduced localized automation, it remains inherently bounded by single-task processing pipelines incapable of handling complex cross-layer dynamics. As a partial remedy, large language models (LLMs) excel at generalized cognitive reasoning, but to a degree they remain detached from the rich modalities of wireless telemetry. As a solution, we unveil Generative Network Intelligence conceptualized via the Perceive-Reason-Act (PERA) paradigm. This paradigm treats the wireless channel and the underlying network states as a continuous, multimodal narrative. By synchronizing the perceptual grounding of Large Wireless AI Models (LWAMs) with the cognitive reasoning of LLMs, PERA heralds the era of native NG intelligence. Crucially, this unified intelligence replaces fragmented, task-specific edge models by an efficient multi-task architecture delivering the real-time control needed for supporting dynamic physical applications while reducing both the complexity and energy dissipation. Moreover, we contrast the structural limitations of traditional ML to generative paradigms, conceive agentic reasoning across a NG protocol stack, and detail a practical three-tier design specifically engineered for the resource-constrained wireless edge. This architectural paradigm serves as a foundational framework for realizing fully autonomous, embodied agentic AI in NG networks. To validate this vision, our case study evaluates link-state classification and beam prediction, demonstrating how grounding wireless telemetry within a cognitive engine delivers the transparent, human-readable rationales required for trusted physical-layer diagnostics and beam control.
\end{abstract}


\section{\bf Introduction} 

\begin{figure*}[t]
    \centering
    \subfloat[6G-enabled business and application verticals 
    \label{fig:intro_use_cases}]{%
        \includegraphics[width=0.49\linewidth]{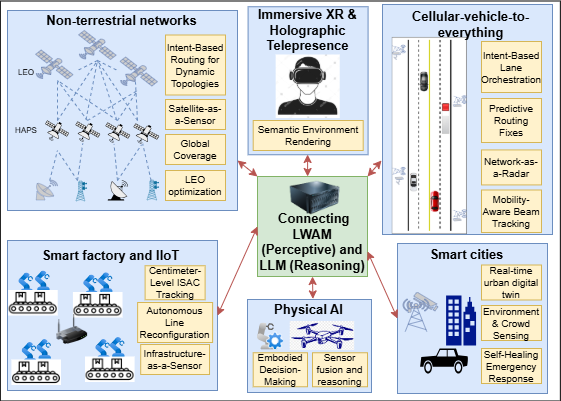}%
    }
    \hfill
    \subfloat[Synchronization of cognitive reasoning and sensory perception across the protocol stack.\label{fig:intro_stack}]{%
        \includegraphics[width=0.49\linewidth]{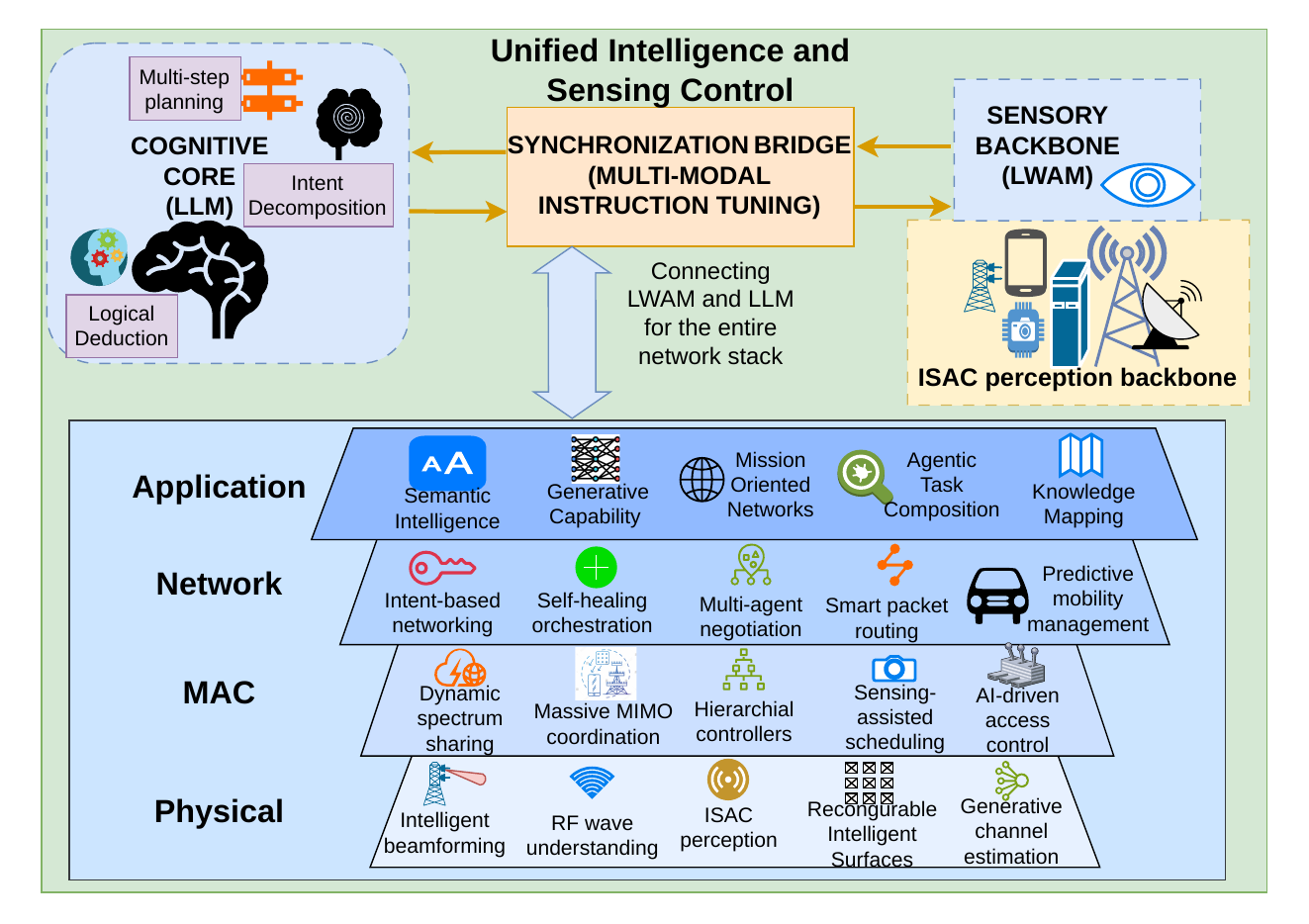}%
    }
    
    \caption{Framework for Embodied \gls{ai}-driven \gls{6g} intelligence: (a) application domains and vertical deployments, and (b) synchronization of cognitive reasoning and sensory perception across the full network stack.}
    \label{fig:framework_overall}
\end{figure*}

The \gls{6g} network is not merely an advanced version of \gls{5g}. It represents a fundamental shift from a deterministic bit-pipe infrastructure to fully autonomous, \gls{ai}-native, agentic system. 
As shown in Fig. \ref{fig:intro_use_cases}, \gls{6g} deployments must simultaneously support massive device connectivity, ultra-reliable low-latency communications, and real-time adaptive intelligence across terrestrial and non-terrestrial domains. Conventional networks cannot meet these requirements~\cite{zhu2025wireless}.
Hence 6G must embrace Embodied AI by grounding its cognitive layer within its physical, spectral, and spatial surroundings \cite{zhu2025wireless}. This enables agentic nodes to self-consciously perceive, reason, and act (the foundational operations of the \gls{pera} paradigm) across the entire layered protocol stack, as depicted in Fig. \ref{fig:intro_stack}, rather than running isolated algorithms in an environmental vacuum.
Consequently, the underlying network fabric must evolve to seamlessly synthesize real-time physical-layer variations with high-level cognitive policies across the entire protocol stack.
Yet, realizing this at scale requires intelligence at the mobile edge, where latency constraints, communication and computing resource limitations, and real-time control requirements force cognitive reasoning to be grounded in physical-layer realities.

Realizing this level of autonomy requires a complete departure from the current status quo of telecommunications \gls{ai}. Today's discriminative and reinforcement learning pipelines lack the agentic capability of handling distributed \gls{ai} tokens or of reasoning over abstract operator intents \cite{ibn, netgpt}. Instead, they operate as isolated, task-specific black boxes that collapse under the cross-layer dynamics and high dimensionality of large-scale \gls{6g} deployments. 
Without a bespoke mechanism to orchestrate these complex networking capabilities, conventional frameworks fail to connect high-level intentional reasoning with the continuous, high-dimensional telemetry of the wireless medium.

Effectively navigating these network inter-dependencies requires the complementary pairing of two distinct foundation model families, yet an inherent architectural barrier divides them.
\glspl{llm} excel at abstract intent decomposition, logical deduction, and multi-step planning \cite{prompt_hao}. However, they remain intrinsically blind to the physical world — unable to natively process continuous wireless features like raw \gls{csi} or \gls{rf} interference patterns. Conversely, \glspl{lwam} act as powerful sensory systems capable of tokenizing raw telemetry and capturing physical invariants \cite{zhu2025wireless, farzanullah2025wireless, alikhani2024large}. However, they lack cognitive reasoning, causal interpretation capabilities, and the capacity to interpret high-level operator mandates. Because existing framework lacks a principled, synchronized interface between these two domains, the \gls{llm} cannot read the continuous wireless medium, and the \gls{lwam} cannot actuate goal-directed policies, leaving a critical integration puzzle.

Motivated by the desire to bridge this dual-model divide, we introduce the Perceive-Reason-Act (\gls{pera}) paradigm. 
At its core, \gls{pera} addresses a fundamental modality mismatch: \glspl{llm} excel at semantic reasoning over text, while \glspl{lwam} excel at feature extraction from raw physical signals. Rather than forcing one of the models to adapt to the other's domain, \gls{pera} establishes a lightweight semantic bridge --- a trainable projector that maps high-dimensional physical features into the embedding space of text tokens. To the \gls{llm}, these ``network pseudo-tokens'' appear as an ordinary sequence of words describing the state of the environment. This alignment transforms \gls{pera} into a unified reasoning engine capable of ingesting network pseudo-tokens at the mobile edge, synthesizing cross-layer network state, and producing actionable outputs grounded in both physical observation and semantic reasoning, thereby enabling embodied, intelligent decision-making across the \gls{6g} protocol stack.

Consolidating this generative interface at the edge yields compelling operational advantages over fragmented, legacy automation. Aligning wireless features with natural language representation can allow network operators to apply advanced prompt engineering—like \gls{icl} and \gls{cot} reasoning \cite{prompt_hao}—directly over models trained on raw wireless data. This eliminates hosting hundreds of task-specific diagnostic heads, drastically reducing edge memory bottlenecks and overhead. Instead, a single foundational core adapts to novel scenarios and anomalies through contextual, natural-language instructions alone, completely bypassing continuous gradient updates or immediate retraining. The proposed \gls{pera} architecture enables these capabilities through unified perception-reasoning integration. Furthermore, by appending recent history to the prompt, the architecture leverages its attention mechanism for real-time predictive decisions while natively delivering human-readable, causal explanations that demystify black-box network decisions.

The remainder of this article provides a comprehensive framework for fully autonomous reasoning and control across the \gls{6g} stack. We first establish a structured, layer-wise categorization of the communication stack, contrasting traditional machine learning limitations against generative paradigms across the physical, network, and application layers. Following this, we detail the practical three-tier design of the \gls{pera} architecture, detailing the mechanics of the universal perceptual encoder, the lightweight modality bridge, and the edge-optimized cognitive engine. Finally, we present an empirical case study on beam prediction and \gls{los} classification using the DeepMIMO dataset to demonstrate how grounding wireless signals within a language framework enables zero-shot environmental adaptability. Furthermore, it delivers the transparent, human-readable \gls{ai} explanations necessary to successfully audit and troubleshoot channel decisions, validating the embodied reasoning paradigm at the core of \gls{pera}.

\section{\bf A Holistic Investigation of the \gls{6g} Stack: Synergizing \glspl{llm} and Large Wireless AI Models} 

Although recent surveys, most notably \cite{zhu2025wireless}, establish a comprehensive taxonomy for \gls{ai}-native \gls{6g}, they focus heavily on adapting \glspl{llm} to wireless tasks via prompting. This approach often overlooks signal-native models, such as the \gls{lwm} \cite{alikhani2024large}, which are pre-trained directly on physical layer representations such as raw \gls{csi} or baseband spectrograms. 
Moreover current literature relies on isolated automation pipelines that lack the capacity to link high-level reasoning with deterministic physical-layer control.
This targeted review identifies these integration gaps to justify the necessity of the unified, edge-deployed architecture proposed in this work. A summary of the works across the network stack is provided in Table \ref{tab:literature_summary}.

\subsection{\bf Physical Layer: Foundation Models and Channel Representation}

In \gls{6g}, the Physical (PHY) layer is envisioned to shift from rigid mathematical models toward universal, generative foundation models that leverage task-agnostic feature extraction for rapid adaptation and out-of-distribution robustness. As foundational pillars, the \gls{lwm} \cite{alikhani2024large} employs \textit{Masked Channel Modeling} to extract universal wireless features, while the multimodal \gls{wmfm} \cite{farzanullah2025wireless} uses contrastive learning to align radio channels with visual imagery for \gls{isac}. By applying \gls{peft} strategies—freezing the backbone and updating only lightweight heads—both models minimize training overhead while outperforming traditional benchmarks in localization and \gls{los}/\gls{nlos} classification. 

To eliminate the preprocessing information loss associated with \gls{csi} or spectrograms, IQFM \cite{iqfm} operates directly on raw multi-antenna IQ waveforms, capturing spatio-temporal signal characteristics via contrastive self-supervised learning. Ultimately, transitioning to these compact, task-agnostic encoders is fundamental to mobile edge intelligence, enabling versatile, real-time physical-layer adaptation on resource-constrained devices.

\subsection{\bf Network Layer: Intent-Based Multi-Agent Orchestration}

The Network layer is evolving toward decentralized cognition via multi-agent systems driven by sophisticated orchestration frameworks. This paradigm shift replaces static, imperative configurations with a goal-oriented operations, enabling the network to autonomously decompose complex operator intents into coordinated actions across distributed edge nodes. 

\begin{itemize}
    \item \textbf{Intent-Driven Management:} A critical advancement in this layer is the utilization of state-space models for intent-to-policy mapping. As detailed in \cite{ibn}, the linear scaling and efficient long-range dependency handling of Mamba-based architectures allow the network to interpret abstract human intents into real-time configuration scripts without the computational overhead of standard transformers. This generative approach ensures that the network agent can maintain state awareness across dynamic \gls{ran} conditions.
    
    \item \textbf{Collaborative \gls{ai}-native Orchestration:} Extending this cognition across the cloud-edge continuum requires heterogeneous \glspl{llm} to operate in coordination, where lightweight edge models enrich user prompts with location-specific context before forwarding them to a more capable cloud model for full response generation \cite{netgpt}. In this framework, a dedicated Computing Plane jointly schedules communication and computation across the hierarchy, balancing the load between edge and cloud based on task complexity rather than replicating full model copies at every node. This hierarchical synergy yields intelligent network management as a byproduct, with edge \glspl{llm} continuously inferring operator intents and predicting traffic demand without any additional orchestration overhead.
    
    \item \textbf{Intent-Driven Dynamic Network Slicing:} To manage \gls{6g} complexity, the framework in \cite{e2e_slicing} adaptively segments physical resources across the \gls{ran}, transport, and core. It utilizes customized \glspl{llm} to iteratively translate operator intents into deployment specifications. Backed by data-driven capacity predictions, this collaborative dialogue allows users to dynamically refine slicing requests, balancing service satisfaction with resource efficiency.
\end{itemize}

\subsection{\bf Application Layer: Generative Reliability and Semantic Intelligence}

The Application layer in \gls{ai}-native \gls{6g} systems marks a fundamental shift toward Generative Network Interfaces, where the network state is treated as a multimodal narrative rather than a set of isolated bit-streams. This transition is anchored by generative diffusion models, which provide a robust framework for network state reconstruction and resilience against extreme throughput variations \cite{genai_diffusion}. By treating signal recovery as an inverse probabilistic process, these models can reconstruct high-fidelity telemetry even from heavily compressed semantic payloads, ensuring generative reliability for mission-critical edge applications. 

Complementing this transition is the integration of edge-based large \gls{ai} models that facilitate cognitive multimodal semantic communication \cite{semanticedge}. These frameworks utilize multimodal tokenizers to eliminate bit redundancy, transmitting only the core semantic meaning of a payload to surpass classical Shannon limits. However, the non-deterministic nature of these agentic systems necessitates rigorous verification. To address this, the \gls{6g}-Bench framework provides an open, standardized benchmark for evaluating semantic communication and network-level reasoning \cite{6g_bench}. By assessing \glspl{fm} across dozens of complex network reasoning tasks, \gls{6g}-Bench validates the integrity of agentic intent parsing and multi-agent coordination, providing the empirical rigor required to transition from discriminative \gls{ai} to reasoning-centric \gls{6g} orchestration.

\subsection{\bf Prompt Engineering techniques in the wireless domain}

Integrating \glspl{llm} into wireless networks via prompt engineering—such as \gls{icl} and \gls{cot}—facilitates resource-efficient management by bypassing the computational demands of fine-tuning \cite{prompt_hao}. At the physical layer, the \textbf{RF-GPT} framework maps IQ waveforms to visual spectrograms, identifying technologies like \gls{5g} and LTE with over 97\% accuracy while providing grounded explanations \cite{rfgpt}. Similarly, applying \gls{cot} reasoning to traffic prediction enhances forecast stability by decomposing temporal dependencies, outperforming standard baselines by up to 22.41\% in $R^2$-score \cite{CoT_mahdi}. These developments signal a shift toward unified \gls{6g} interfaces where \glspl{llm} natively query and explain complex spectral dynamics in real-time.

\begin{table*}[t] 
\centering
\caption{Layer-wise Literature Categorization: Features, Challenges, and Applications}
\label{tab:literature_summary}
\renewcommand{\arraystretch}{1.3}
\begin{tabular}{|p{3.0cm}|p{4.2cm}|p{4.2cm}|p{4.2cm}|}
\hline

\textbf{Category} & \textbf{Main Features} & \textbf{Challenges} & \textbf{Applications} \\
\hline

\textbf{Physical Layer Foundation Models} \cite{alikhani2024large, farzanullah2025wireless, iqfm}
&
\begin{itemize}[leftmargin=*,nosep]
    \item Task-agnostic transformer encoders using Masked Channel Modeling and contrastive self-supervised learning.
    \item Multimodal alignment of radio channel coefficients with visual imagery for \gls{isac}.
    \item Direct operation on raw IQ signals, bypassing preprocessed inputs such as CSI.
\end{itemize}
&
\begin{itemize}[leftmargin=*,nosep]
    \item High computational overhead for real-time edge inference.
    \item Generalization across heterogeneous propagation environments with minimal retraining.
    \item Capturing joint temporal and spatial characteristics from raw waveforms.
\end{itemize}
&
\begin{itemize}[leftmargin=*,nosep]
    \item \gls{los}/\gls{nlos} classification and channel-based localization.
    \item Zero-shot multi-task sensing without manual labels.
    \item Task-agnostic embeddings for versatile physical-layer edge intelligence.
\end{itemize}
\\
\hline

\textbf{Intent-Based Network Orchestration} \cite{ibn, netgpt, e2e_slicing}
&
\begin{itemize}[leftmargin=*,nosep]
    \item Mamba-based state-space models for efficient, low-overhead intent-to-policy mapping.
    \item Heterogeneous \glspl{llm} distributed across the cloud-edge continuum with a dedicated Computing Plane.
    \item Iterative negotiation with residual capacity prediction for end-to-end network slicing.
\end{itemize}
&
\begin{itemize}[leftmargin=*,nosep]
    \item Maintaining transactional integrity and safe rollbacks under dynamic \gls{ran} conditions.
    \item Coordination of multi-agent negotiations across \gls{ran}, transport, and core domains.
    \item Balancing edge-side prompt enhancement with cloud-side generation under resource constraints.
\end{itemize}
&
\begin{itemize}[leftmargin=*,nosep]
    \item Converting abstract operator intents into real-time \gls{ran} configuration scripts.
    \item Self-healing and autonomous network orchestration.
    \item Dynamic end-to-end network slicing with user-collaborative refinement.
\end{itemize}
\\
\hline

\textbf{Generative and Semantic Application Layer} \cite{genai_diffusion, semanticedge, 6g_bench}
&
\begin{itemize}[leftmargin=*,nosep]
    \item Generative diffusion models treating signal recovery as an inverse probabilistic process.
    \item Multimodal tokenizers to eliminate bit redundancy and transmit only core semantic meaning.
    \item Open standardized benchmark evaluating foundation models across complex network reasoning tasks.
\end{itemize}
&
\begin{itemize}[leftmargin=*,nosep]
    \item Reconstructing high-fidelity telemetry from heavily compressed semantic payloads.
    \item Non-deterministic agentic behavior requiring rigorous empirical verification.
    \item Defining evaluation metrics for multi-agent coordination and agentic intent parsing.
\end{itemize}
&
\begin{itemize}[leftmargin=*,nosep]
    \item Resilience against extreme throughput variations for mission-critical edge applications.
    \item Surpassing classical Shannon limits via cognitive semantic communication.
    \item Validation of agentic intent parsing and multi-agent coordination performance.
\end{itemize}
\\
\hline

\textbf{Prompt Engineering for Wireless Systems} \cite{prompt_hao, rfgpt, CoT_mahdi}
&
\begin{itemize}[leftmargin=*,nosep]
    \item \gls{icl}, \gls{cot}, and self-refinement strategies enabling complex optimization via forward passes only.
    \item Multimodal grounding mapping IQ waveforms to time-frequency spectrograms via STFT.
    \item Plan-based CoT pipeline generating intermediate rationales for temporal dependency decomposition.
\end{itemize}
&
\begin{itemize}[leftmargin=*,nosep]
    \item Avoiding fine-tuning overhead on power-limited and resource-constrained edge devices.
    \item Handling spectral diversity across heterogeneous wireless technologies.
    \item Ensuring forecast stability across non-stationary mobile traffic patterns.
\end{itemize}
&
\begin{itemize}[leftmargin=*,nosep]
    \item Network optimization and prediction without model retraining.
    \item \gls{rf} technology identification (\gls{5g} NR, LTE, Bluetooth) exceeding 97\% accuracy \cite{rfgpt}.
    \item Mobile traffic forecasting with up to 22.41\% $R^2$ improvement over \gls{icl} baselines \cite{CoT_mahdi}.
\end{itemize}
\\
\hline

\end{tabular}
\end{table*}

\subsection{\bf Synergetic Empowerment: The Unified Organism}
These trajectories converge into \textit{Synergetic Empowerment}, transforming the \gls{6g} network into a unified architecture where \glspl{llm} provide the cognition and \gls{lwam} serve as the sensory body. Fully realizing this embodiment requires a seamless bridge between high-level reasoning and wireless actuation. The following section proposes a framework where instruction tuning acts as the critical connective tissue, utilizing prompt-based \gls{ai} to align agentic intent with the multi-modal complexities of the wireless medium. 

\section{\bf The \gls{pera} Architecture: Perceive-Reason-Act for Edge Intelligence} \label{sec3} 

\begin{figure*}
    \centering
    \includegraphics[width=1\linewidth]{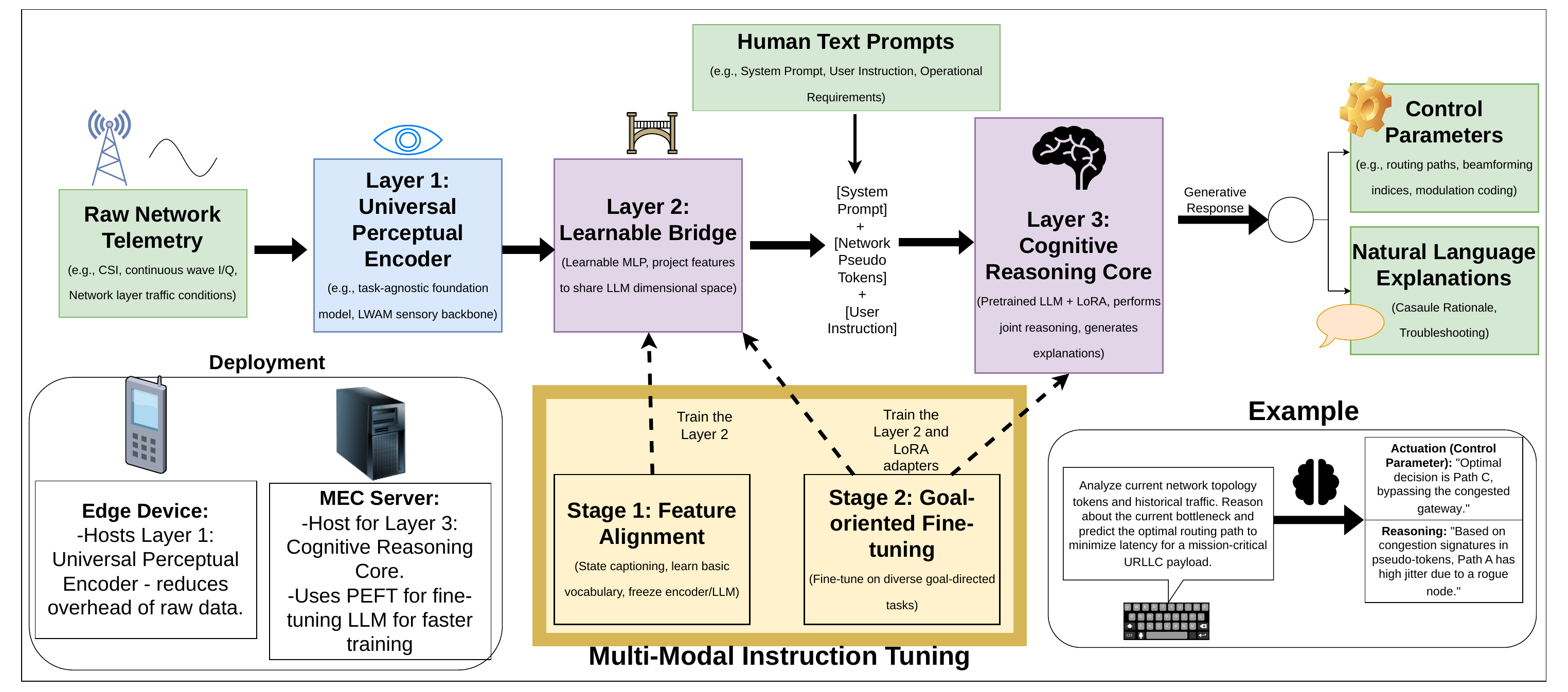}
    \caption{Overview of the \gls{pera} framework's unification strategy, depicting the two-stage training protocol: Stage 1 (Feature Alignment) and Stage 2 (Goal-oriented Fine-tuning) to bridge the gap between physical \gls{rf} signals and semantic reasoning.  }
    \label{fig:pera}
\end{figure*}

Realizing Embodied \gls{ai} for Mobile Edge General Intelligence requires moving beyond fragmented, single-task models. The diversity of mobile networks---from dynamic physical layer signal variations to application layer traffic orchestration---demands a unified framework. We therefore propose the \gls{pera} architecture.

\gls{pera} bridges raw, multi-dimensional network states across any protocol stack layer with the reasoning capabilities of \glspl{llm}, replacing discrete task-specific architectures with a generalized, cross-layer framework explicitly designed for the memory, computational, and latency constraints of heterogeneous Mobile Edge General Intelligence environments.

\subsection{\bf The High-Level Philosophy: From Discriminative \gls{ai} to Generative Network Interfaces}

The \gls{pera} architecture represents a fundamental shift from discriminative \gls{ai} toward a ``Generative Network Interface.'' Traditional \gls{ai} in telecommunications relies on discriminative models that act as opaque black boxes, outputting isolated labels without contextual reasoning. Operating an intelligent network edge therefore requires an \gls{mec} server to host hundreds of disparate models, creating memory bottlenecks, orchestration overhead, and poor generalization to novel scenarios.

\gls{pera} replaces this monolithic approach with a generative paradigm in which the wireless channel and network state are treated as a \textit{Multimodal Narrative}. An \gls{llm} serves as the ``Cognitive Core,'' supported by specialized foundation models as ``Perceptual Sensory Systems.'' Rather than emitting context-less labels, the architecture maps raw network state into a semantic latent space, conditioning the \gls{llm} to output a joint probability distribution alongside a natural language reasoning chain. A single \gls{fm} at the edge can thus ingest raw data from any protocol stack layer, reason over dynamic network intents, and produce context-aware execution strategies.

\subsection{\bf The Three-Tier Architectural Stack}

The \gls{pera} architecture, shown in Fig. \ref{fig:pera}, utilizes a three-tier modular design to optimize resource efficiency by distributing tasks between the \gls{mec} server and edge devices. This separation of actions ensures that heavy cognitive tasks rely on the \gls{mec} server, while light-weight sensory encoding harnesses edge devices, such as IoT gateways or user equipment. 

\subsubsection{\bf Layer 1: The Universal Perceptual Encoder}
The foundational layer consists of domain-specific, yet task-agnostic, Perceptual Encoders. 

\textbf{Function:} The primary role of this layer is to ingest raw, high-dimensional, and often non-linear network data. Because we design the \gls{pera} framework to be completely generic, the input data modality is highly flexible. It may involve physical layer I/Q continuous waves, MAC layer scheduling matrices, or network-layer traffic flow topologies.
Since Layer 1 remains frozen at the device, deployment on resource-constrained hardware is achieved through standard model-based compression techniques (quantization, distillation, pruning), making \glspl{lwam} practical for inference.

\textbf{Design:} The Perceptual Encoder is typically realized through a Transformer-based architecture that utilizes spatial-temporal patching strategies. Through large-scale self-supervised pre-training, this encoder learns to tokenize the raw input, capturing deep physical invariants and structural correlations while ignoring localized noise. 

\textbf{Output:} The encoder produces a sequence of high-dimensional latent vectors. These vectors represent the underlying structural features of the network environment abstracted entirely from the raw telemetry.

\subsubsection{\bf Layer 2: The Modality Bridge (Semantic Projector)}

\textbf{Function:} The fundamental challenge is an inherent language barrier: \glspl{llm} ``speak'' the language of human semantics (text), whereas the Perceptual Encoder ``speaks'' the language of physics and network dynamics.
The Modality Bridge translates between these domains by mapping wireless features into the \gls{llm}'s text embedding space.

\textbf{Mechanism:} This layer is implemented via a lightweight adapter having significantly fewer parameters than the \gls{llm} and Perceptual Encoder. This efficiency makes it computationally suitable for edge deployment without sacrificing alignment quality.

\textbf{Role:} It projects the perceptual vectors into a sequence of ``Network Pseudo-Tokens.'' These pseudo-tokens are mapped to share the exact dimensional space as the \gls{llm}'s standard text embeddings. Consequently, to the \gls{llm}, the raw network telemetry simply appears as a sequence of words describing the state of the environment in an unfamiliar, yet mathematically justified text.

\subsubsection{\bf Layer 3: The Edge-Optimized Cognitive Reasoning Core} ~\\

\textbf{Function:} The apex of the \gls{pera} architecture is the Cognitive Reasoning Core, instantiated by a pre-trained \gls{llm}. Its objective is to perform joint reasoning over human-readable text instructions and the network pseudo-tokens generated by Layer 2.

\textbf{Input:} The \gls{llm} ingests a concatenated sequence of multimodal inputs structured as: $\text{[System Prompt]} + \text{[Network Pseudo-Tokens]} + \text{[User/System Instruction]}$. The instructions can be provided by operators, derived from network 
policies, or generated by \gls{llm} agents.

\textbf{Output:} The model outputs a comprehensive generative response that includes both deterministic control parameters (e.g., precise routing paths, specific modulation and coding schemes, and beamforming indices) alongside natural language explanations detailing its logical reasoning process.

To ensure that this architecture is feasible within the strict memory limits of a Mobile Edge node, the Cognitive Core leverages \gls{peft}. Specifically, the foundational weights of the \gls{llm} remain entirely frozen. Adaptation to specific network tasks is achieved using \gls{lora} modules injected into the attention layers. By updating only a minuscule fraction of the model's total parameters, \gls{pera} enables dynamic, real-time task switching; the edge server can swap out lightweight \gls{lora} adapters in milliseconds without needing to unload and reload massive foundation models in its limited memory.

\subsection{\bf The Unification Strategy: \gls{mmit}}

The innovation of the \gls{pera} architecture lies not just in its individual structural components, but in the methodology used to fuse them. In order to bind the frozen perceptual encoder, the lightweight Modality Bridge, and the \gls{lora}-adapted \gls{llm} core, we use \gls{mmit} adopted from \cite{mmit}. This two-stage protocol is designed to force the \gls{llm} to comprehend raw network telemetry as a foundational context for complex problem-solving.
Unlike monolithic instruction tuning, \gls{mmit}'s separation of semantic alignment and task-specific reasoning enables explicit grounding in physical reality, essential for network safety and reliability.

\subsubsection{\bf Stage 1: Representation Alignment}
Before an \gls{llm} can effectively optimize a network, it must first inherently understand what the mathematical representations of network data actually mean. The goal of Stage 1 is to teach the \gls{llm} the fundamental vocabulary of the telecommunications stack. 

\textbf{Task:} During this phase, the system performs a task analogous to ``State Captioning.'' A massive dataset of paired network states and text descriptions is utilized. For example, given a latent representation of a complex multipath fading environment or a highly congested routing table, the target output is a precise descriptive text (e.g., \textit{``This sequence represents a high-mobility, non-line-of-sight signal path with severe Doppler shifting''}).

\textbf{Mechanism:} Only the Modality Bridge (Layer 2) is trained during this stage. The Perceptual Encoder and the \gls{llm} remain entirely frozen. The loss function applied is the standard Causal Language Modeling loss over the predicted description tokens. This phase acts as a translation dictionary, mechanically mapping the geometric properties of the network latent space into the semantic topology of the \gls{llm}'s word embeddings.

\subsubsection{\bf Stage 2: Goal-Oriented Instruction Tuning}
Once the semantic alignment is established, the system must learn to execute complex, goal-directed edge-intelligence tasks. 

\textbf{Goal:} To solve specific networking, communication, and orchestration challenges using natural language reasoning and instructions.

\textbf{Mechanism:} In this stage, the Modality Bridge remains trainable, and specific \gls{lora} adapters are activated within the \gls{llm}. The model is fine-tuned on a diverse dataset of instruction-response pairs covering multiple domains, such as \gls{isac}, predictive handover optimization, and dynamic bandwidth allocation.

\textbf{Instruction Format:} A training sample in this phase takes the following format:
\begin{itemize}[leftmargin=*]
    \item \textbf{Instruction:} ``Analyze the channel state and determine  the optimal beam group.''

    \item \textbf{Response:} ``Energy peaks at the center with high L2 norm agreement. Beam Group 5 dominates. Select Beam Group 5 for optimal signal alignment.''
\end{itemize}
Through this extensive Instruction Tuning, the architecture transitions the \gls{llm} from a passive text generator into an active, embodied agent capable of parsing complex human intents and executing highly logical, deterministic actions within the mobile edge environment.

\subsection{\bf Key Advantages for Mobile Edge General Intelligence}

Transitioning to the unified, generative \gls{pera} architecture optimizes the wireless edge by consolidating diverse operations—such as localization and congestion control—into a single foundational model, drastically reducing edge server memory and overhead. Driven by \gls{llm} reasoning, its zero-shot adaptability allows the network to fluidly handle novel environments and anomalies through contextual instructions without requiring immediate retraining. Furthermore, \gls{pera} solves the telecom ``black-box'' dilemma by providing native explainability, generating human-readable, causal rationales for critical actions like bandwidth reallocation to streamline human troubleshooting. 
These capabilities can translate to measurable performance improvements in classification accuracy, inference latency, and network throughput.
Finally, by appending recent history to the prompt, the model leverages its attention mechanism for in-context predictive tracking, dynamically forecasting near-future states while completely bypassing continuous gradient updates.

\section{\bf Case Study: \gls{pera} Implementation via Physical Layer Instruction Tuning} 

To validate the theoretical framework established in the preceding sections, this section presents a comprehensive empirical evaluation of the \gls{pera} architecture. We evaluate the performance of this unified architecture within a simulated environment, focusing on the efficacy of the two-stage \gls{mmit} protocol in translating high-dimensional network telemetry into actionable, reasoning-centric responses.

\subsection{\bf Experimental Setup and Dataset Curation}

The experimental setup utilizes spatial channel characteristics and \gls{los}/\gls{nlos} metadata derived from the DeepMIMO dataset \cite{DeepMIMO}. We employ the \gls{lwm} \cite{alikhani2024large} as our Universal Perceptual Encoder to ingest complex \gls{csi} and high-dimensional \gls{rf} waveforms. The \gls{lwm} functions as a universal feature extractor, tokenizing these raw \gls{csi} matrices into structured latent representations that capture deep physical invariants of the wireless environment. To bridge the gap between these physical representations and linguistic reasoning, we constructed an instruction-tuning dataset by mapping the tokenized channels into custom-curated supervised dialogue pairs. Each instance bundles a natural-language task instruction with the \gls{lwm}-generated observation to produce a response defining the link condition and optimal beam index.

\subsection{\bf Implementation and Training Protocol}

The implementation follows a two-phase strategy designed to establish a shared "dialect" between the wireless and language domains. 

In the first phase, \textbf{Representation Alignment}, a lightweight modality bridge projects the wireless global summary token into the high-dimensional embedding space of the \gls{llm}. This bridge, implemented as a two-layer \gls{mlp} projector with a hidden width of 1024 and GELU activation, is optimized using the AdamW optimizer with a learning rate of $10^{-4}$ and a batch size of 4. During this stage, the \gls{llm} backbone remains frozen while the bridge learns to map latent wireless features to qualitative text descriptions.

In the second phase, \textbf{Goal-Oriented Instruction Tuning}, the model performs joint reasoning over the instruction prompt and the wireless token. To preserve physical fidelity and prevent generative "hallucinations", we incorporated auxiliary pathways to predict link state and coarse beam groupings directly from the embeddings.
By grounding the language output in physical labels, these auxiliary tasks prevent the model from generating inaccurate or physically inconsistent explanations.
The beam-group head utilizes a residual 1-D \gls{cnn} stem having a width of 256 and five residual blocks, while the \gls{los} head consists of a two-layer \gls{mlp}. Physical fidelity is enforced through a combined objective function appropriately weighting language modeling, \gls{los} classification, and beam-grouping by a ratio of $0.3:1.0:0.5$, respectively. 

For this instruction tuning, the \gls{llm} is wrapped with \gls{lora} adapters on the query, key, value, and output projections (rank 16, $\alpha=32$). Optimization proceeds through a three-stage curriculum of two epochs each, with AdamW learning rates decaying from $10^{-3}$ to $10^{-4}$. This staged approach ensures the interface remains faithful to physical labels while optimizing for the multi-task reasoning capabilities of the \gls{llm}.
For the \gls{llm} in the cognitive reasoning core, we use the Qwen-3 Instruct Model, as its advanced intent decomposition and multi-step planning capabilities allow the network to translate abstract human mandates into deterministic configurations while maintaining the high-fidelity reasoning required for explainable \gls{6g} intelligence.

During the training process, we utilized five scenarios from the DeepMIMO dataset, including San Diego, Santa Clara, Fortworth, Indianapolis, and Oklahoma. To evaluate generalization, the inference was conducted on six novel scenarios. For the beam prediction task, the model was configured to identify 16 distinct beam groups. 

To rigorously assess the performance of the \gls{pera} architecture, two benchmarks were developed for comparison. The first benchmark employs a standard \gls{cnn} trained on raw \gls{csi} data, serving as a baseline for traditional discriminative approaches. The second benchmark trains identical diagnostic heads directly from the \gls{lwm} representations to isolate the impact of \gls{lwm} driven feature extraction. 
 
\subsection{\bf Experimental Results} 

\begin{figure}
    \centering
    \includegraphics[width=1\linewidth]{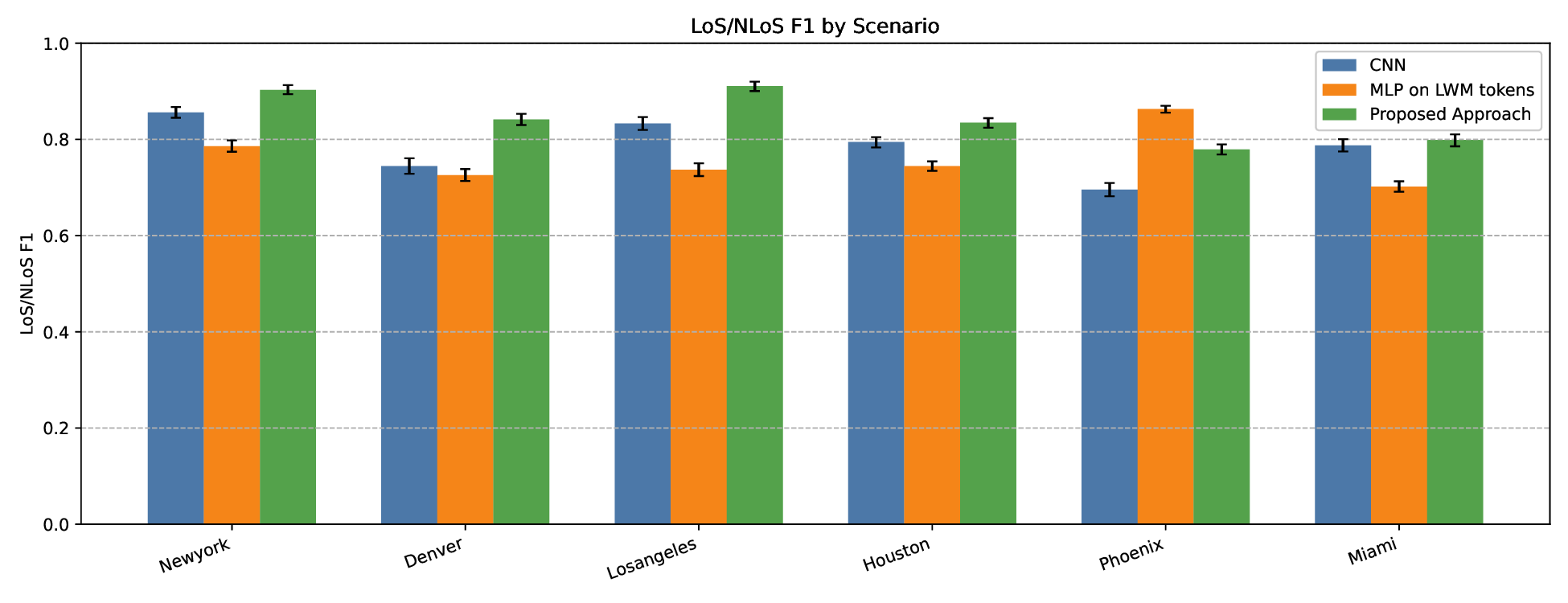}
    \caption{Comparison of F1 score for \gls{los}/\gls{nlos} classification.}
    \label{fig:losf1}
\end{figure}

\begin{figure}
    \centering
    \includegraphics[width=1\linewidth]{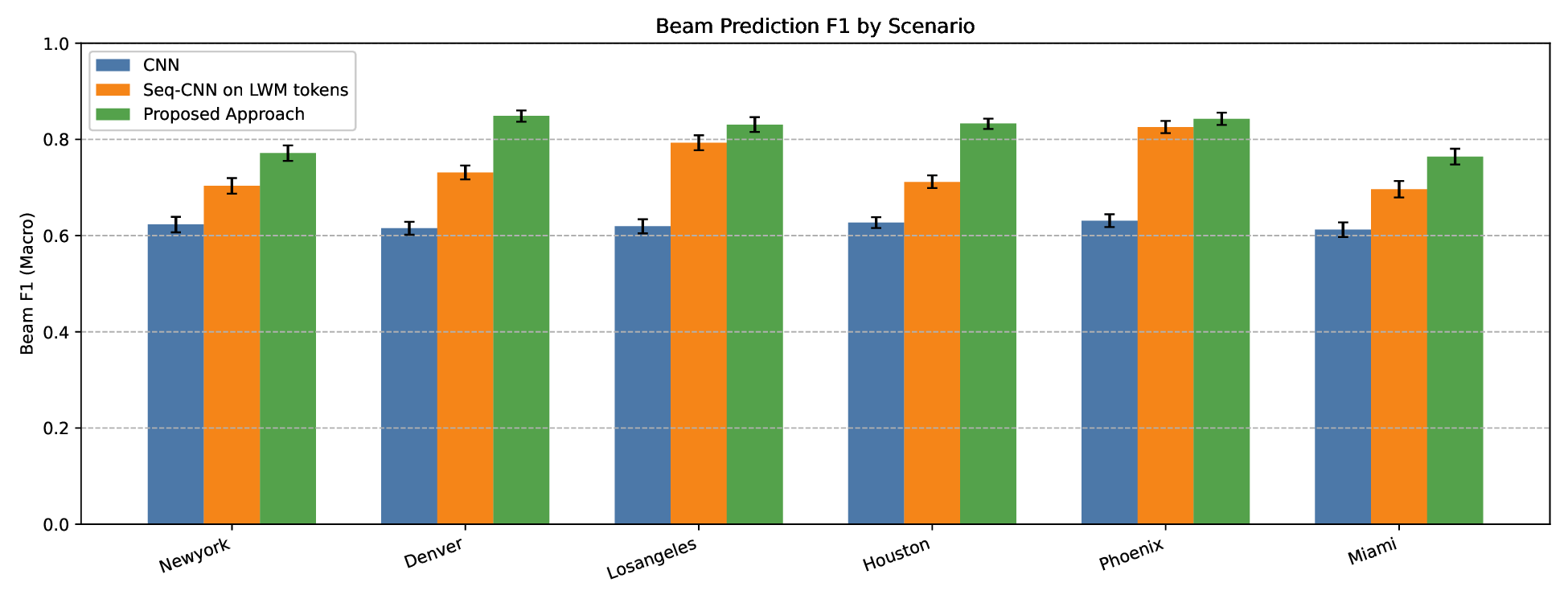}
    \caption{Comparison of F1 score for beam prediction.}
    \label{fig:beamf1}
\end{figure}

The results of our empirical evaluations, illustrated in Fig. \ref{fig:losf1} and Fig. \ref{fig:beamf1}, demonstrate the superior performance of the \gls{pera} architecture across diverse novel scenarios. Fig. \ref{fig:losf1} compares the F1 scores for \gls{los} and \gls{nlos} classification, where the proposed approach consistently outperforms the raw \gls{csi}-trained \gls{cnn} by $7.73\%$ and the \gls{mlp} (\gls{lwm}) baseline by $11.77\%$ in the environments considered. 

Similarly, Fig. \ref{fig:beamf1} highlights the framework's efficacy in beam prediction, showing a substantial performance gain of $31.17\%$ over the baseline \gls{cnn} and a $9.90\%$ improvement over the Seq-CNN (\gls{lwm}). These findings indicate that by intrinsically integrating the cognitive reasoning of the \gls{llm} with the perceptual grounding of the \gls{lwm}, the system achieves more agile environmental adaptability and confident decision-making than traditional discriminative models. Furthermore, the stability of these results across spatially distinct scenarios, such as Houston and Miami, validates the architecture's adaptability to time-variant propagation conditions in the \gls{6g} mobile ecosystem.

\begin{figure}
    \centering
    \includegraphics[width=1\linewidth]{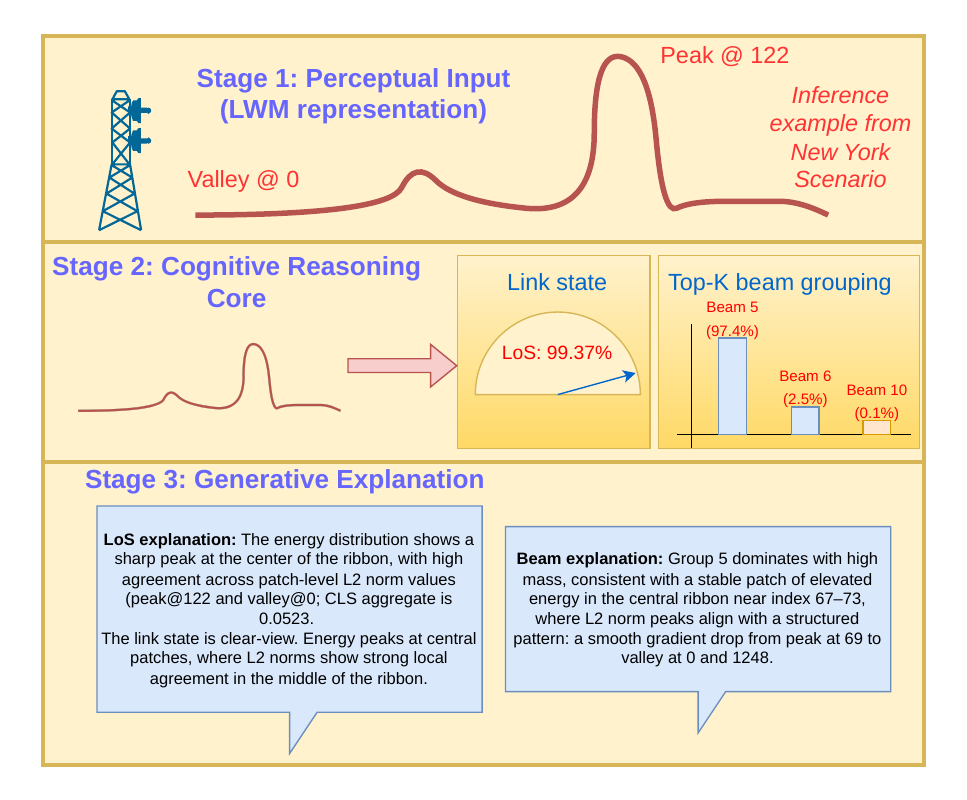}
    \caption{Demonstration of the \gls{pera} architecture pipeline during inference in an urban scenario. Stage 1 illustrates the perceptual energy distribution extracted from the channel tokens. Stage 2 shows the cognitive reasoning core classifying the link state as \gls{los} ($99.37\%$) and identifying the optimal beam group ($97.4\%$). Stage 3 presents the generative explanations that ground the network's decisions in the observed physical invariants.}
    \label{fig:inference}
\end{figure}

The end-to-end inference pipeline of the \gls{pera} architecture is demonstrated in Fig. \ref{fig:inference}, using a test sample. In Stage 1, the perceptual input extracted from the channel tokens reveals a specific energy distribution. In Stage 2, the Cognitive Reasoning Core processes these representations to classify the link state as \gls{los} with 99.37\% confidence and identifies Beam 5 as the optimal beam group with 97.4\% probability. Finally, Stage 3 generates explanations that ground these decisions in observed physical invariants. For instance, the system provides a causal rationale for the \gls{los} classification by identifying high agreement across patch-level L2 norm values, while the beam selection is justified by the stable patch of elevated energy detected in the central ribbon.

This transparent reasoning mechanism enables the realization of Explainable \gls{ai}, eliminating the ``black-box'' limitation of traditional models by providing a human-readable, causal rationale that is critical for streamlining auditing and troubleshooting processes in complex network environments.

\section{\bf Conclusion} 

A generative network interface architecture realizing "Agentified Embodied" \gls{ai} for Mobile Edge General Intelligence was conceived for \gls{6g}. By synchronizing an \gls{lwam} sensory backbone with an \gls{llm} cognitive engine via a two-stage \gls{mmit} protocol, \gls{pera} overcomes traditional discriminative \gls{ai} limitations: task-specificity, black-box opacity, and poor generalization. Our cross-layer survey demonstrates this convergence across the protocol stack: physical-layer foundation models offer task-agnostic sensory grounding, intent-based frameworks enable autonomous, agentic orchestration, and application-layer diffusion and semantic communication systems push past classical Shannon limits. Together, they establish a unified network architecture fusing cognition and perception.
Empirical validation on the DeepMIMO dataset confirms that \gls{pera} consistently outperforms discriminative baselines in previously unseen scenarios, achieving $+9.90\%$ in beam prediction and $+7.73\%$ in \gls{los}/\gls{nlos} classification F1 scores. Crucially, these gains are paired with human-readable, causal explanations, directly mitigating the explainability concern impeding operator trust in channel diagnostics and beam control.

\section*{Acknowledgment}

The authors gratefully acknowledge Dusit Niyato, from the Nanyang Technological University, Singapore, for his insightful feedback and detailed review of this manuscript. His critical suggestions significantly strengthened the technical depth and clarity of the presentation.

\bibliographystyle{IEEEtran} 
\bibliography{references}

\vfill

\end{document}